# Record nonlinear conversion efficiency in the production of high spectral purity vacuum ultraviolet laser at 148 nm

SERGEY VASILYEV[1,*], TIAN OOI[2,**], IGOR MOSKALEV[1], MIKE MIROV[1], ANDREY MURAVIEV[3], DMITRII KONNOV[3], VICTOR CHURIKOV[1], VIKTOR SUKHAREV[1], EVGENY GALENIN[1], JACK F. DOYLE[2], CHUANKUN ZHANG[2], KAI LI[2], GEORGIY SERYOGIN[1], DAN PERLOV[1], IGOR SAMARTSEV[1], KONSTANTIN VODOPYANOV[3], JUN YE[2,***]

[1]*IPG Photonics Corporation, 377 Simarano Drive, Marlborough, MA 01752 USA*
[2]*JILA, NIST and University of Colorado, Department of Physics, University of Colorado, Boulder, CO 80309 USA*
[3]*CREOL, The College of Optics and Photonics, 4304 Scorpius Street Orlando, FL 32816 USA*
**svasilyev@ipgphotonics.com, **tian.ooi@colorado.edu, ***ye@jila.colorado.edu*

**Abstract:** Coherent vacuum-ultraviolet (VUV) lasers are indispensable for precision measurement, quantum optics, and materials science. Recent high-resolution spectroscopy of the Th-229 nuclear clock transition near 148 nm highlights the urgent demand for intense, narrow-linewidth VUV lasers for advancing metrology and testing fundamental physics. However, existing VUV generation schemes typically require enhancement cavities [C. Zhang *et al.*, Opt. Lett. 47, 5591–5594 (2022)], atomic resonances [Q. Xiao *et al.*, Nature 650, 852-856 (2026)], or random quasi-phase-matched nonlinear crystals [V. Lal *et al.*, Optica 12, 1971–1974 (2025)]. Here, we demonstrate a VUV frequency comb via cascaded frequency doubling of a 2400 nm Cr:ZnS comb to its 16th harmonic in nonlinear crystals. The final stage employs a bulk-grown, spatially uniform quasi-phase matched (QPM) crystal developed by IPG, combining VUV transparency, high $\chi^{(2)}$ nonlinearity, and power scalability. Using this QPM crystal we generate a VUV frequency comb with 40 μW average power (1 nW per mode at 80 MHz mode spacing) with a conversion efficiency order of magnitude higher than other known methods. These results establish a scalable route to compact VUV sources via direct frequency doubling, opening a path toward a robust continuous-wave nuclear clock laser.

## 1. Introduction

The first resonant laser excitation of a nuclear transition in Th-229 was recently reported in 2024 using nanosecond pulsed lasers with 20 GHz resolution [1, 2]. The resolution and precision have been improved by more than 5 orders of magnitude using a high spectral-purity and frequency-stabilized VUV optical frequency comb [3], demonstrating a reproducible frequency ratio between this nuclear transition and a Sr optical atomic clock [4, 5]. These experiments marked the culmination of years of effort by generations of scientists [6-10], which was initiated by a prediction of a low-energy isomeric state in the Th-229 nucleus by Kroger and Reich in 1976 [11].

The development of Th-229 nuclear clocks based on this transition promises a revolution in metrology [12, 13]. The Th-229 transition's high frequency, intrinsically narrow linewidth, and insensitivity to external perturbations compared to electronic transitions make it an ideal candidate for a high accuracy optical clock [14,15]. Importantly, the nuclear transition is preserved in a solid-state crystalline environment [16-18], allowing for orders of magnitude higher number of emitters compared to current atomic clocks. Furthermore, coherent control of quantum states in the Th-229 nuclear system opens new avenues in quantum optics [19]. It has also been proposed to use the exceptional properties of the Th-229 isomer in searches for new physics, e.g. tests of Einstein's equivalence principle and time variation of fundamental constants [20-25].

Availability of VUV laser sources at 148 nm is key to the development of Th-229 nuclear clocks, motivating new advances in laser science. This task is challenging as the requirements imposed on this new type of lasers are stringent. First, the laser's linewidth must be ultranarrow, ideally below 1 Hz, to approach the linewidth of the nuclear transition. Second, for initial spectroscopic studies, a nW-level power is sufficient. However, practical nuclear clock operation, nonlinear spectroscopy, and quantum optics all require continuous wave (cw) lasers with low intensity noise and with μW to mW power level. Finally, field applications such as future transportable nuclear clocks, require compact and reliable lasers with low complexity and power consumption.

All current implementations of 148 nm sources used for high resolution spectroscopy of the Th-229 are based on nonlinear up-conversion of existing lasers. The examples are illustrated in Fig. 1 and include: (a) a comb source based on non-perturbative high-harmonic generation (HHG) of a cavity-enhanced ultrafast Yb:fiber laser system [26], (b) a cw source based on four-wave mixing (FWM) of two Ti:Sa lasers in cadmium vapor [27], and (c) a cw source based on second harmonic generation (SHG) in randomly quasi-phase-matched strontium tetraborate crystal (RQPM-SBO) [28]. The parameters of these sources are summarized in Table 1.

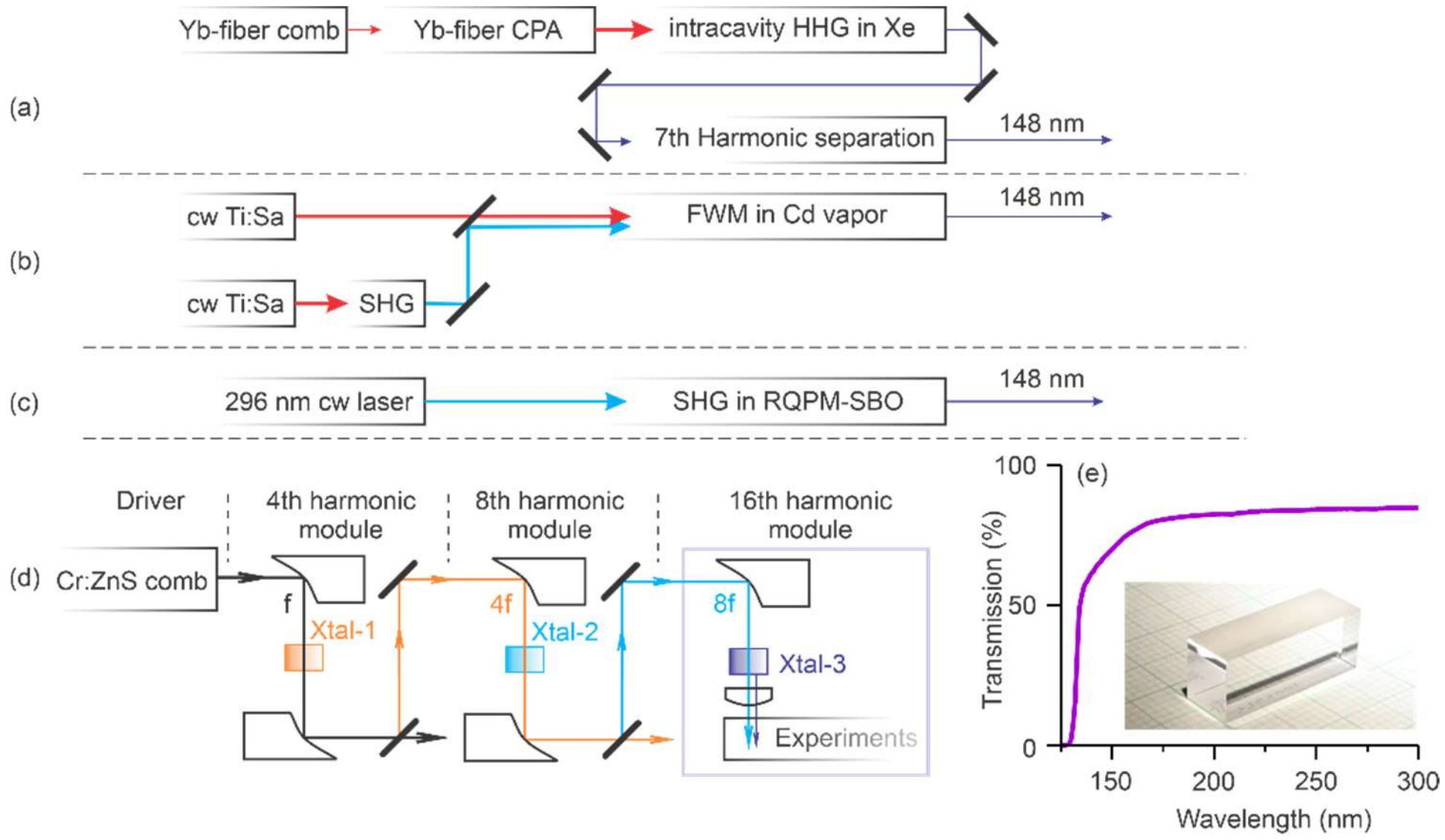

Fig. 1. Comparison between existing approaches for highly coherent VUV light generation and the record efficiency three wave mixing process via quasi phase matching reported in this work. (a) high harmonic generation; (b) four-wave mixing; (c) second harmonic generation; (d) 16$^{th}$ harmonic of the Cr:ZnS comb via QPM three wave mixing; (e) VUV transmission of a 10 mm thick sample of new VUV QPM nonlinear material used. An insert shows a photo of a 20 mm long element.

**Table 1. Parameters of laser sources for high resolution Th-229 nuclear spectroscopy**

| Type | Parameters of the driving laser [a] | | | | Nonlinear process | VUV power [b], nW | Ref. |
|---|---|---|---|---|---|---|---|
| | λ, nm | P, W | Δν, kHz | Δτ, fs | | | |
| comb | 1036 | 50 | ~ 1 | 150 | HHG | 10 | [26] |
| cw | 710 + 375 | 4 + 2 | $10^{-3}$ | -//- | FWM | 100 | [27] |
| cw | 296 | ~ 0.4 | 100 [c] | -//- | TWM | 1 | [28] |
| comb | 2400 | 2.5 | ~ 7 [c] | <24 | TWM | 1 | This work |

[a] Wavelength (λ), power (P), linewidth (Δν), pulse width (Δτ); [b] Available for the Th-229 spectroscopy; [c] Without active stabilization

In terms of power scaling, the HHG-based sources are hindered by the requirement of extreme peak laser intensity, the inherently low efficiency of high-order nonlinear optical processes, as well as the broad spectral span consisting of multiple harmonic orders. Assessment of various constraints such as saturation effects, thermal effects, and phase stability suggests that output power of cw sources based on FWM in atomic vapors is potentially scalable to 0.1 – 10 mW [29], with a record power of 10 μW recently demonstrated and used for nuclear absorption-based clock [30]. However, the high-temperature vapor cells and multi-watt driving lasers required for this approach may complicate implementation of compact, power efficient, field-deployable systems.

Compared to HHG and FWM approaches, three-wave mixing (TWM) in crystalline materials provides the highest conversion efficiencies and, hence, offers a simpler and more power scalable route to VUV sources for Th-229 spectroscopy. In this respect, work by Lal et al. [28] about a 148 nm source based on TWM in randomly quasi-phase-matched SBO is promising and has also been used for a nuclear absorption-based clock [31, 32]. All other things being equal, the yield of a TWM process can be increased by several orders of magnitude if one uses a birefringent phase-matched or quasi-phase-matched (QPM) material instead of an RQPM material with poorly defined microstructure [33]. Yet, identification and fabrication of nonlinear crystals that have VUV transparency and can be phase-matched for the target wavelength of 148 nm is challenging. For example, TWM below 200 nm has been demonstrated in $KBe_2BO_3F_2$ [34], $LiB_3O_5$ [35], and $NH_4B_4O_6F$ [36], however these crystals are not transparent at 148 nm. $BaMgF_4$ has a transparency cutoff at 125 nm but TWM has only been reported down to 355 nm, likely limited by phase-matching and material considerations [37,38].

In this work, we report a proprietary crystalline material developed by IPG Photonics Corp that uniquely combines VUV transparency extending to 130 nm, high $\chi^{(2)}$ nonlinearity, and QPM capability. Further, the parameters of the QPM structure of this material can be optimized for different TWM processes for any target output wavelength down to 130 nm. Here we report the evaluation of this new material's SHG efficiency in the 148 nm regime for the application in Th-229 nuclear spectroscopy. The VUV light is generated as the 16th harmonic of a commercial optical frequency comb at the fundamental wavelength of 2.4 μm. The obtained results provide new estimates on achievable parameters of pulsed and continuous wave 148 nm lasers for future nuclear clocks, atomic and molecular spectroscopy, and quantum optics.

## 2. Experimental setup and results

The schematic of the VUV laser source is illustrated in Fig. 1(d) and its main parameters are summarized in Table 2. For the driving laser, we used an optical frequency comb based on ultrafast polycrystalline Cr:ZnS laser technology [39]. For the VUV light generation, the fundamental comb is frequency-doubled two times in the same PPLN crystal with two consecutive QPM gratings. The first grating is optimized for f → 2f SHG and the second grating is optimized for 2f → 4f SHG process. Then the obtained 4f comb is focused into a BBO crystal optimized for 4f → 8f SHG process. The 8f comb is then coupled to IPG Photonics Corp proprietary crystalline material, which is indicated in Fig. 1(d) as Xtal-3. The QPM structure of Xtal-3 was optimized for 8f (1012 THz) → 16f (2025 THz) SHG process.

**Table 2. Parameters of the driving laser and nonlinear crystals.**

| Driver | Xtal-1 | Xtal-2 | Xtal-3 |
|---|---|---|---|
| Laser medium: Cr:ZnS[a] | PPLN | BBO | IPG Proprietary Material |
| Power: $P > 5W$[b] | 0.5 mm thick | 0.5 mm thick | 2 mm thick |
| Wavelength: $\lambda = 2.4$ μm | uncoated | AR/AR coated | uncoated |
| Pulse width: $\tau < 24$ fs | 2 QPM gratings | $\theta = 36.4°$ | 1 QPM grating |
| Mode spacing: $f_{rep} = 80$ MHz | f → 2f → 4f | $4f_{(o)}+4f_{(o)}\rightarrow 8f_{(e)}$ | 8f → 16f |

[a] Post-grown thermal diffusion doped Polycrystalline Cr:ZnS; [b] During the experiments average power was limited to 2.5 W to prevent the PPLN crystal from damage.

Figure 2 summarizes parameters of the obtained 4f signal in the visible (VIS) and 8f signal in the ultraviolet B (UVB) spectral ranges. For the power measurements in Fig. 2 (a), the power of the Cr:ZnS MOPA (master oscillator power amplifier) was varied by adjusting the power amplifier setpoint, which resulted in variations of the spatial, spectral and temporal distributions of driving pulses. These variations partially explain the saturation of 4f and 8f signals as the input power increases. The autocorrelation of 8f pulse is shown in Fig. 2 (b), which reveals the pulse duration of ~ 200 fs. Observed astigmatism of the 8f beam shown in Fig. 2(c) suggests somewhat imperfect alignment of the parabolic mirrors and spatial walk-off in the BBO crystal. At the maximum driving power of 2.5 W, the obtained 4f and 8f power levels are 283 mW and 96 mW, respectively, corresponding to 11% and 4% conversion efficiency of driving pulses to the VIS and to the UVB.

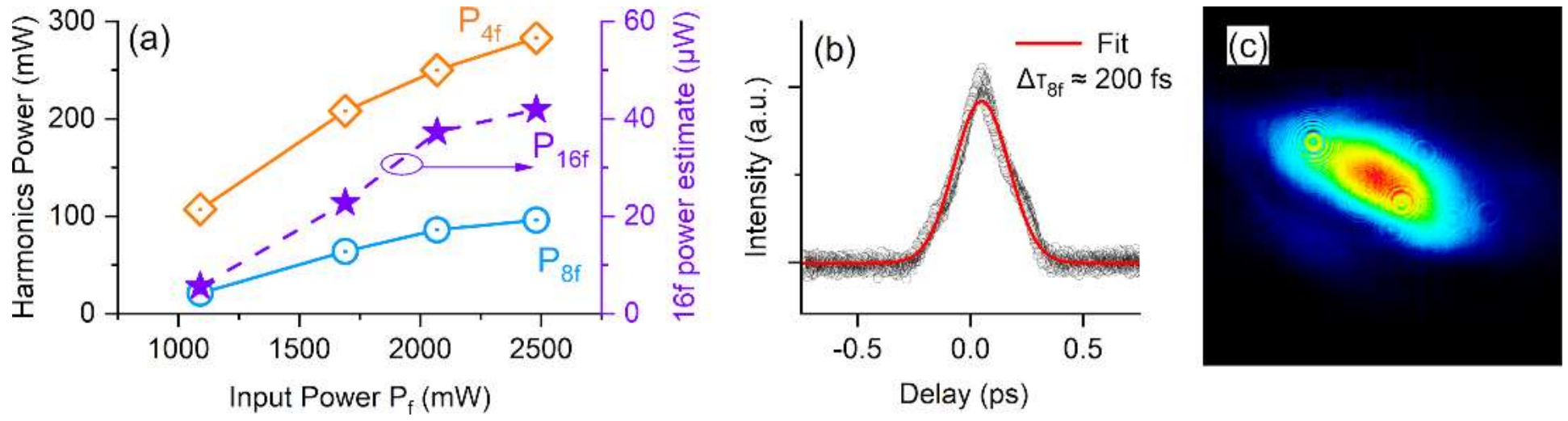

Fig. 2. Preparation of 8f light before the final 8f→ 16f frequency conversion. (a) left vertical axis, measured power of 4f and 8f signals vs driving laser power; (a) right vertical axis, estimated power of the VUV 16f vs driving laser power (see text). (b) second-order autocorrelation of 8f pulses measured with an APE pulseCheck autocorrelator (circle); Gaussian fit of autocorrelation (red line). (c) Beam profile of 8f signal measured with a DataRay WinCamD-LCM.

The obtained 8f comb was coupled to an airtight chamber and focused onto the Xtal-3 sample using a 25.4 mm focal length parabolic mirror. At the maximum 2.5 W power of the driving laser, we measured 60 mW of 8f power at the input of the Xtal-3 sample, corresponding to an estimated 3.75 kW peak power. The optical signal at the output of the Xtal-3 was collimated by a VUV-grade $CaF_2$ lens (eSource Optics) with a focal length ~ 40 mm. In most experiments, spectral components at the output of the Xtal-3 sample were separated by a VUV-grade $CaF_2$ prism (eSource Optics) and a VUV dichroic mirror (Layertec). In some experiments the prism was replaced by an Edmund Optics metal-coated VUV mirror. During experiments on the VUV generation, the airtight chamber was filled with argon gas at 1 – 10 mbar. We confirmed the presence of the VUV component and made background corrections by comparing the measurements in argon and in ambient air, which absorbs only 16f and transmits all lower harmonics.

The experiments on the VUV light generation consisted of three parts, as illustrated in Fig. 3(a-c). First, we assessed the VUV generation in a small signal regime by attenuating the 8f input signal to 0.5 mW average power and 30 W peak power. As shown in Fig. 3(a), the VUV signal was measured with a photomultiplier tube (PMT, Hamamatsu R6835). A 50 μm slit was installed at the front of the PMT's input window to shield it from the background light. As shown in Fig. 3(d), the measured dependence of the VUV signal on the input 8f signal is quadratic, which indicates a TWM process. We then scanned the VUV beam across the slit to estimate its size (~ 0.5 mm) and used the PMT specifications to estimate the VUV power (0.4 nW after the slit and ~ 4 nW for the whole beam). This corresponds to a projection of 40 μW VUV power at the full power of 8f signal.

To measure the VUV power with high 8f power, the PMT was replaced with a VUV enhanced Si photodiode (Opto Diode AXUV100G) to avoid saturation, as shown in Fig. 3(b). To attenuate the background signal, to which the photodiode is quite sensitive, a VUV bandpass filter was installed at the PD's hood. The photocurrent was measured while the chamber was filled with argon and with air, and in this way evaluated the optical power of the VUV component at the PD (5.7 μW max). The data on the losses of optical components was then used to estimate the VUV power at the output of the Xtal-3 sample (50 μW max).

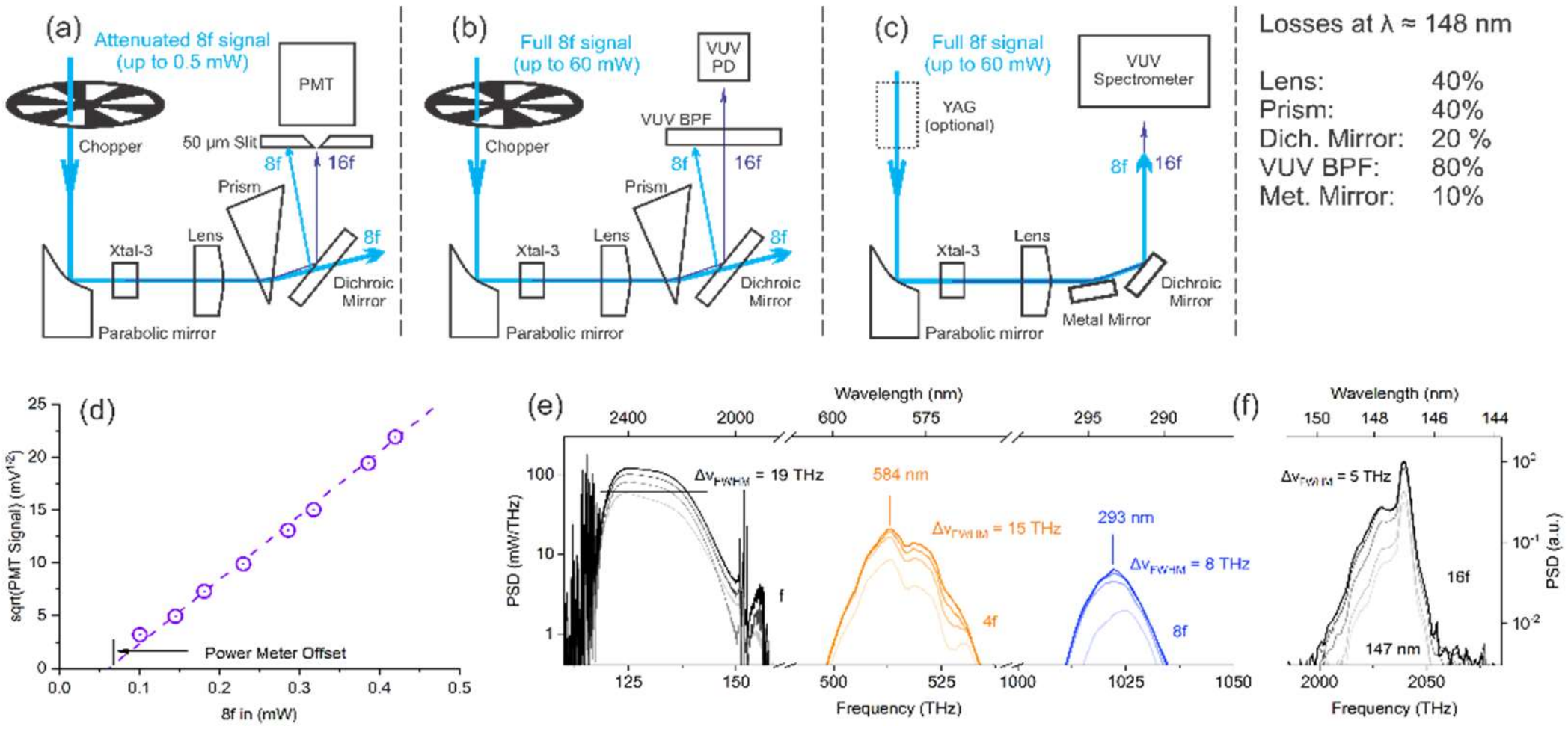

Fig. 3. (a, b, c) Setup for the VUV light characterization (see text) and approximate losses of the optical components at the wavelength of 148 nm. (d) Measured dependence of the square root of the PMT signal vs. average power on input 8f signal. (e) Spectra f, 4f and 8f signals measured with a Thorlabs OSA207 (MIR range), an Ocean Optics USB4000 (VIS-UVB ranges). Thick lines correspond to the maximum power of input driving pulses; thin lines correspond to the intermediate power levels of input driving pulses. (f) Spectrum of the 16f signal measured with an HP Spectroscopy easyLIGHT VUV spectrometer. Power spectral density (PSD) of the spectral distributions in part (e) was normalized using the power values from Fig. 2(a). PSD of the spectral distributions in part (f) was normalized to unity. Top horizontal axes in parts (e) and (f) show wavelength in reciprocal scale.

The measured spectra of the driving laser and its 4$^{th}$ and 8$^{th}$ optical harmonics are shown in Fig. 3 (e). The up-conversion via TWM process resulted in spectral narrowing, which can be expected for the sub-3-cycle driving pulses. The measured 200 fs duration of 8$^{th}$ harmonic is well above their Fourier transform limit (about 60 fs and 40 fs for Gaussian pulse and sech$^2$ pulse, respectively) indicating the pulse is stretched. To characterize the VUV spectrum and confirm the VUV power, we connected an aberration-corrected normal-incidence VUV spectrometer with ~ 0.1 nm resolution equipped with a VUV enhanced CCD sensor. The VUV spectra is illustrated in Fig. 3(f), which indicates the VUV pulse is centered at ~ 147 nm and the bandwidth is ~ 5 THz (FWHM), further narrowed due to the quasi-phase-matching bandwidth. We scanned the VUV beam across the spectrometer's input slit to estimate its size (~ 0.44 mm). Integrating the spectral power and accounting for the sensor's efficiency and the losses of optical components, the VUV power at the output of the Xtal-3 sample was estimated to be 42 μW.

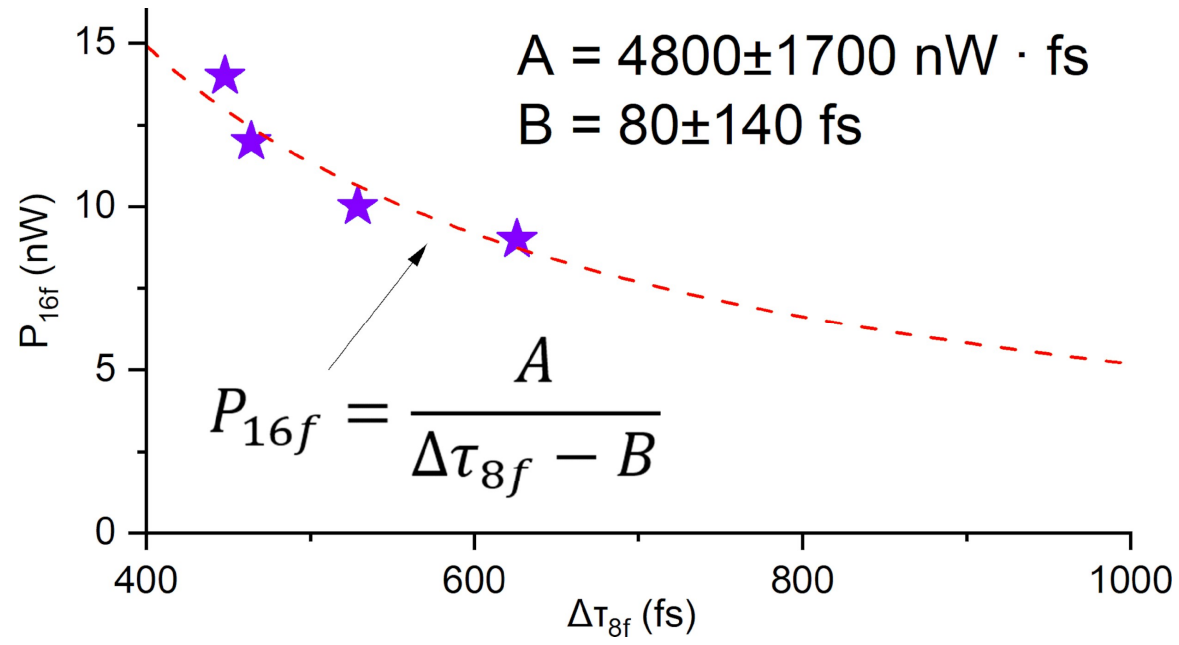

Fig. 4. XUV signal power ($P_{16f}$) vs 8f signal pulse duration ($\Delta\tau_{16f}$) at 2.5 mW 8f signal averaged power: experimental measurements (symbols); data fit by a hyperbola (dashed curve).

The power of a cw VUV laser, which is the most relevant optical local oscillator for a Th-229 nuclear clock, will be explored in future work. An initial estimate can be inferred from the effect of pulse stretching on 16f power. The input 8f pulses were stretched using a set of rods and plates made of undoped YAG. The obtained dependence of the output VUV power on the duration of input pulses is shown in Fig. 4. Assuming the average VUV power generated from the SHG process is inversely proportional to the 8f pulse duration, we fit the experimental data with a hyperbolic function $A/(\Delta\tau_{8f} - B)$. A simple extrapolation of the VUV power to cw regime $P_{16f}^{cw} = P_{16}^{cw}\,(\Delta\tau_{8f} \rightarrow 1/f_{rep})$ yields 0.38(0.14) pW for 2.5 mW input 8f power. This would scale to around 60 nW of cw VUV power at 1 W cw input power. The corresponding VUV SHG efficiency would be ~ $6\times10^{-8}$, which is an order of magnitude higher than the conversion efficiency reported in [28]. Practically, other factors such as focusing, thermal effects, walk-off, and chirp will complicate this extrapolation, which we only use as an order of magnitude estimate.

## 3. Conclusions and outlook

We demonstrate the up-conversion of an optical frequency comb at the fundamental wavelength 2.4 μm to the VUV via the f → 2f → 4f → 8f →16f TWM process in a chain of nonlinear crystals. The last crystal in the chain is the QPM crystalline material with high VUV transparency and high $\chi^{(2)}$ nonlinearity, and the QPM structure are optimized for the 8f = 1012 THz → 16f = 2025 THz SHG.

The measurements of the VUV spectra and power in the small signal regime confirm that this QPM crystalline sample indeed outputs VUV light at the 16$^{th}$ harmonic of the fundamental comb. We estimated the 16th harmonic power at the output of the sample using a photomultiplier, photodiode and an enhanced CCD sensor. The obtained values are in good agreement: 40 μW for photomultiplier, 50 μW for photodiode and 42 μW for CCD.

Thus, we achieved 1 PHz → 2 PHz SHG with $7\times10^{-4}$ efficiency for the pulsed input signal with 60 mW average power and 3.75 kW peak power. To the best of our knowledge, this is the highest conversion efficiency reported to date for nonlinear processes involving high spectral purity VUV light. The generated VUV signal is a frequency comb with 80 MHz mode spacing that can be readily phase-locked to any standard optical frequency time-base and used for high resolution spectroscopy, including dual comb spectroscopy.

The power spectral density of the VUV comb (1 nW per comb mode) is a little lower than that obtained using HHG technique (10 nW per comb mode at production) [26]. However, we achieved this power level – already relevant for the spectroscopy of the Th-229 nuclear clock transition – with several orders of magnitude higher conversion efficiency and greatly reduced complexity of the whole system.

The extrapolations from the pulsed to continuous-wave regime yield, conservatively, 60 nW of the VUV power at 1 W cw sub-harmonic input power for the 1 PHz → 2 PHz SHG in the 2 mm thick NLE, which is comparable to the cw power achieved using FWM technique [27] and 60 times higher than the RQPM [28]. However, it is important to mention that the VUV power level in cw regime can be straightforwardly increased by a factor of 5-to-10 using practically feasible 10-to-20 mm long NLEs, and even higher VUV power levels can be reached using enhancement cavities, thus providing a practical path to mW-class sources at the VUV wavelength 148 nm.

These results highlight the main advantages of VUV light generation via TWM process in now available (albeit proprietary) QPM materials, namely high efficiency in a solid-state medium. It enables implementation of power-scalable, compact, simple and reliable VUV laser sources for high resolution spectroscopy, quantum optics, and transportable nuclear clocks.

## 4. Back matter

**Funding.** Work at JILA is funded by the DARPA (HR0011-25-3-0140), the Army Research Office (W911NF-25-1-0244), DOE Quantum Center of Quantum System Accelerator, and the National Institute of Standards and Technology. Work at CREOL was funded by US Office of Naval Research (ONR) award N00014–18–1-2176; US Air Force Office of Scientific Research (AFOSR) awards FA9550–23–1-0126 and FA9550–24–1-0196; and U.S. Department of Energy, awards DE-SC0012704 and DE-FG02-09ER16021.

**Acknowledgment.** S.V., I.M., and M.M. thank the scientific and engineering teams at IPG Photonics for their support. Certain commercial equipment and materials are identified in this paper to foster understanding. Such identification does not imply recommendation or endorsement by the National Institute of Standards and Technology, nor does it imply that the materials or equipment identified are necessarily the best available for the purpose.

**Disclosures.** The authors declare no conflicts of interest.

**Data availability.** Data underlying the results presented in this paper are not publicly available at this time but may be obtained from the authors upon reasonable request.